\documentclass[12pt, preprint]{aastex}

\received{2005 February 24}
\begin{document}


\title{GALAXY-QUASAR CORRELATIONS BETWEEN APM GALAXIES AND HAMBURG-ESO QSOS}

\author{Joshua G. Nollenberg\altaffilmark{1}, Liliya L.R. Williams}
\affil{Astronomy Department, School of Physics and Astronomy,
University of Minnesota, 116 Church St. S.E.,
Minneapolis, MN 55455}
\email{jgnollenberg@stthomas.edu, llrw@astro.umn.edu}

\altaffiltext{1}{Current address: Department of Physics, University of St. Thomas,
OWS 153, 2115 Summit Ave., St. Paul, MN 55105 }

\begin{abstract}

We detect angular galaxy-QSO cross-correlations between the APM Galaxy Catalogue
and a preliminary release (consisting of roughly half of the anticipated final catalogue)
of the Hamburg-ESO Catalogue of Bright QSOs as a function of source QSO
redshift using multiple cross-correlation estimators.  
Each of the estimators yield very similar
results, implying that the APM catalogue and the Hamburg-ESO survey are
both fair samples of the respective true galaxy and QSO populations.  
Though the signal matches the expectations of 
gravitational lensing qualitatively, the strength of the measured 
cross-correlation signal 
is significantly greater than the CDM models of lensing by 
large scale structure would suggest.  This same disagreement between 
models and observation 
has been found in several earlier studies.  We estimate our confidence in the 
correlation detections versus redshift by generating 1000 random realizations 
of the Hamburg-ESO  QSO survey: We detect physical associations between galaxies
and low-redshift QSOs at $99\%
$ confidence and detect lensing associations at roughly $95\%
$ confidence for QSOs with redshifts between 0.6 and 1.  Control
cross-correlations between Galactic stars and QSOs show no signal. Finally, the
overdensities (underdensities) of galaxies near QSO positions relative to 
those lying roughly
$135 - 150\arcmin$ away are uncorrelated with differences in Galactic extinction
between the two regions, implying that Galactic dust is not significantly
affecting the QSO sample. 

\end{abstract}


\section{Introduction}
\label{intro}

\citet{T80}, noting the `apparent extreme evolution' of the QSO
luminosity function, first indicated along with \citet{C81} that 
magnification bias due to gravitational lensing could create statistical 
associations between foreground galaxies and high-redshift QSOs. A number 
of studies have since examined the phenomenon.

Magnification bias arises when foreground gravitational lenses (i.e. galaxy
clusters or large scale
structure) magnify 
the source planes of background objects (QSOs).  This source plane magnification
results in two effects:  1) Individual sources  
span a larger area on the sky than they would if a lens were not present.  
Because gravitational lensing conserves flux, this results in a net 
brightening of the source, which can lead to the inclusion of otherwise
undetectable objects in a
flux-limited sample.  2)  The number density of the sources 
on the sky is decreased because of the same stretching of the source 
plane by the lens.  

Assuming a flux-limited sample of QSOs, 
depending on the slope of the magnitude number counts of a sample of 
selected sources, $\alpha_m$, the combination of the two effects 
produces a statistical association, $\omega_{GQ}(\theta)$, or alternatively
an overdensity (or underdensity), 
$q(\theta)$, which is dependent on the magnification, $\mu(\theta)$:
\begin{equation}\label{lens} q(\theta) - 1 = \omega_{GQ}(\theta) = 
\mu(\theta)^{2.5 \cdot \alpha_m - 1} - 1.\end{equation} If $\alpha_{m} > 0.4$,
positive correlations will be observed, whereas negative
correlations will occur if $\alpha_{m} < 0.4$.  Should $\alpha_{m}$ equal
the critical value of $0.4$, 
no correlations will be detected regardless of lens magnification.
Finally, the amount of magnification depends on the projected surface mass density
of the lens and on the relative distances of sources 
and lenses from the observer.  

There are, however, complications.  
For instance, if dust were located inside galaxy clusters, then QSOs lying on lines of
sight near foreground clusters would tend to be more heavily obscured than those
not lying behind clusters.  This would result in a net decrease in a measured
galaxy-QSO cross-correlation.  A wide variety of techniques have been used to search for dust in galaxy
clusters spanning their range in cluster richness.  For a more complete
description of past work in this area, please refer to the introduction 
section of \citet{NWM03}. Early studies performed on the basis of visual 
inspection \citep{Z62,KL69,dVC72} tended to report rather large amounts of 
extinction (up to $A_V \sim 0.4$ magnitudes) in clusters such as Coma and Virgo.
Smaller (though occasionally mutually inconsistent) estimates have been 
reported by later
studies \citep{w93,H92,B99,AM98,VD95,AJ93,RM92,VV89,M95,DRM90,F93,b90,NWM03},
including upper limits of $E(B-V) \approx 0\fm02$ and $A_R = 0\fm025$ on
reddening and extinction by large scale distributions of dust
within medium-richness APM clusters, respectively \citep{NWM03}.  Low levels 
of dust in galaxy 
clusters such as this imply that cluster dust does not significantly
contribute to galaxy-QSO associations, though the matter has not yet been
entirely resolved because the possibility exists that dust may reside in small
clumps which could have escaped detection using the \citet{NWM03}
technique.  

Galactic dust is also a thorny issue because its presence could result in positive
correlations between galaxies and QSOs or between galaxy clusters and QSOs. 
Due to the fact that Galactic dust extinguishes light from galaxies and QSOs alike,
the observer would tend to see QSOs and galaxies that lie in the same unobscured 
parts of the sky, yielding a positive statistical association. 
Its distribution is also quite filamentary and this structure
continues to scales smaller than are sampled by either the $\sim 6\farcm1$ pixel size of the
\citet{SFD98} extinction map or the lower resolution \citet{BH82} map. 
High-contrast filaments of dust could therefore obscure individual sources, and
while statistical association measurements generally sample scales much larger
than this, Galactic dust could in principle leave its imprint by introducing
power on a wide variety of scales that are dependent on the topology of the
dust distribution itself.  

This paper is organized as follows:  After a discussion of prior results in
Section~\ref{es}, we will describe the galaxy and
QSO data that we use in Section~\ref{data}.  
Then we will discuss our methods for estimating
galaxy-QSO cross-correlationsin Section~\ref{proc}.  Following this, 
we will discuss the results of
our analysis, including measurements of angular cross-correlations, 
integrated cross-correlations vs. redshift, confidence tests and comparisons 
with models through the remainder of Section~\ref{proc}.  Finally, we summarize
our results in Section~\ref{disc}.

\section{Earlier Studies}
\label{es}

Many past observations qualitatively support the statistical lensing picture.
Characteristics of their respective studies, including galaxy and QSO
catalogues, brightness ranges, QSO redshift ranges, as well as the observed
association, can be found in Table 1.  The `Detected' column indicates the 
type of statistically significant cross-correlations that were
detected.  `$+$', `$0$' and `$-$' refer to {\it positive-}, {\it no-}, and 
{\it negative-}correlations, respectively. The studies are divided into three
groups from the top of the table to bottom, respectively:  {\it Top:} Measurements of
galaxy overdensities using incomplete QSO samples.  {\it Middle:} Galaxy-QSO
angular cross-correlations using complete sets of optically-selected  (or in one
case, UV-selected) QSOs. {\it Bottom:} Cross-correlations involving 
complete catalogues of radio-selected QSOs.  {\it Note:} \citet{RM92} and \citet{M95} are
also mentioned in the introductory text, though they did not specifically
measure overdensities or cross-correlations using a single range of QSO
magnitudes and are as a result not listed on this table.

Several early studies, including those found in the top portion of Table 1:
 \citet{T86, F88, F90, HLF90, HRV91, BS94}, and \citet{d92}; 
were conducted using heterogeneous or incomplete QSO
samples, often involving portions of the \citet{HB80} or \citet{VV89} QSO
catalogues, and searched for associations of foreground galaxies within some
angular scale around background QSOs relative to the field average.  Not all of
these studies necessarily attributed measured overdensities with gravitational
lensing.  However, even those studies that did not attribute them to 
lensing do, in retrospect, qualitatively agree
with the magnification bias picture.  

While using optically-selected QSO catalogues, at least four studies found
positive correlations, as expected on the basis of their specific QSO magnitude
cuts.  \citet{w88} detected positive correlations between QSOs and
galaxies on arcsecond scales 
and indicated that the foreground lensing mass must be much greater
than that associated with the visible galaxies themselves.  \citet{RWH94}
observed an overdensity of LBQS QSOs with $1.4 < z < 2.2$ and $B \leq 18\fm5$
within $6.6h^{-1}$ Mpc of Zwicky clusters (which have a typical redshift around 0.2) with
$4.7\sigma$ significance.  Another study, by \citet{WI98} explored statistical
correlations between LBQS QSOs and APM galaxies with a median redshift of $z
\approx 0.25$.  Significant positive correlations were found on angular 
scales $< 100\arcmin$ , using a sample of
QSOs at redshifts $z > 1$ with magnitudes $B_J \leq 18\fm0$.  Finally, \citet{G03} 
used Sloan Digital Sky Survey Early Data Release galaxies and QSOs (albeit a heterogenous sample)
and found cross-correlations on scales $< 10\arcmin$.  Each of these
studies measured cross-correlations that were a few times larger than 
what would be expected from CDM large scale structure lensing models.

Even though the amplitudes of correlations are typically much greater than
expected from models, the signal is most probably due to lensing because of the
unique signature of lensing:  Positive, negative and null correlations have 
been detected using QSO samples with the
expected corresponding brightness cuts.  This may hold true for some studies
that were not originally looking into gravitational lensing as well.
\citet{BFS88}, for example, detected an $30\%
$ deficiency in the number of $B < 20\fm0$ high-redshift UVX-selected QSOs 
at separations within $4\arcmin$ of foreground clusters of galaxies.  Citing
extinction due to dust as the simplest explanation, they ascribed an average
extinction of $A_V \approx 0\fm2$ to their set of clusters.  However, an
anti-correlation is expected within the magnification bias picture for QSOs with
this limiting magnitude due to the resulting shallow slope, $\alpha_m$. 
In fact, \citet{CS99} investigated the \citet{BFS88} anti-correlation under the
assumption that statistical lensing was the cause and found
no discrepancy in interpretation between it and the positive correlation measured by \citet{WI98}.
Similarly, \citet{RM92} measured a $\sim 25\%
$ dearth of high-redshift QSOs within $60\arcmin$ of nearby Abell Clusters and
explained that it could be accounted for by the presence of $A_B = 0\fm15$ of
extinction within the clusters.  Later analysis by \citet{M95} tested the
presence of dust by looking for differences between the colors of QSOs that lay
within $1\degr$ of Abell clusters with those outside of that separation.  No
difference was found and tight constraints on reddening were given: $E(B - V)
\leq 0\fm06$ to $90\%
$ confidence.  Just as in the case of \citet{BFS88}, the anti-correlations 
that were measured in the \citet{RM92} work can likely be explained in terms of
magnification bias.  \citet{RM92} show that the average QSO magnitude in their
sample increases with decreasing QSO-cluster separation.  In fact, the average
magnitude reaches $V \sim 19$ for separations within $10\arcmin$.  Because the
{\it limiting} magnitude is important in magnification bias, while neither
individual QSO magnitudes nor limiting magnitudes of the QSO sample are given,
it is quite likely that the slope, $\alpha_m$ decreases with decreasing
separation (and increasing limiting magnitude, inferred from the increase in the
average magnitude).  This would render a result consistent with weak lensing. 

A natural way to test for the influence of dust (Galactic or extragalactic)
would be to compare
cross-correlations involving optically-selected QSOs with those that utilize
radio-selected ones.  Many of the older studies listed above do use
radio-selected QSOs.  In addition, several studies have incorporated
radio-selected QSOs in order to determine whether extinction is important. 
Comparing radio- and optically-selected subsets of QSOs from the Parkes and LBQS
catalogues, respectively,  with similar redshift distributions 
$(z \geq 0.3 )$, \citet{BM97} noted a $\approx 99\%
$ significant excess of radio QSOs within $10\arcmin$ of COSMOS/UKST galaxies.  The Parks
excess is anti-correlated with the LBQS $(B_J < 20\fm5)$ sample at high
confidence.  And while magnification bias can account for part of the
difference, they argue that dust could account for the rest if the extinction
were present in clusters at a level consistent with \citet{M95}.  
\citet{NW00} measured correlations between 1 Jy QSOs and APM galaxies on 
degree scales and found
positive correlations that are inconsistent with a Galactic dust scenario while 
\citet{BSM01} used two complete samples of QSOs and found a similar result with
the added caveat that the introduction of strong lensing could help to describe
the discrepancy between lensing observations and models on smaller angular
scales.

In general, prior work in galaxy-QSO associations are at least
qualitatively consistent with expectations from magnification bias, though
some measurements yielded results that were significantly greater than predicted
by CDM large scale structure lensing models.  
However, inspection of Table 1 demonstrates that those studies using faint
optical QSOs tend to have negative correlations while those that used bright
QSOs found positive correlations.  This is entirely consistent with
Equation~\ref{lens}.  
  
\section{Data}
\label{data}

\subsection{APM Catalogue}
\label{apm}

Our galaxy data originated from the UKST SES $B_{J}$ and $R$
sky surveys, the photometry of which have been digitized using the 
Automated Plate Machine (APM) at the University of Cambridge \citep{I96}.  
The APM galaxy catalogue, which has limiting magnitudes of $B_{J} \approx 
22.5$ and $R \approx 21$, is complete to magnitudes of $B_{J} \approx 20.5$
and $R \approx 19.5$ with some plate-to-plate variation due to uncertainties 
in absolute photometric calibration and variations in sky background
for each field.  We have not sought to
correct for the variation in photometric calibration, rather we adopt an
analysis strategy that should remove its effect, as we will describe in 
Section~\ref{est}. We culled objects that were detected and classified as 
galaxies in the $R$ band with magnitudes in the range $18.5 \leq R \leq 20$
 from 209 $5\fdg8 \times 5\fdg8$ UKST sky survey fields 
lying at declinations below $+2.5\degr$. 

\begin{figure}

\plottwo{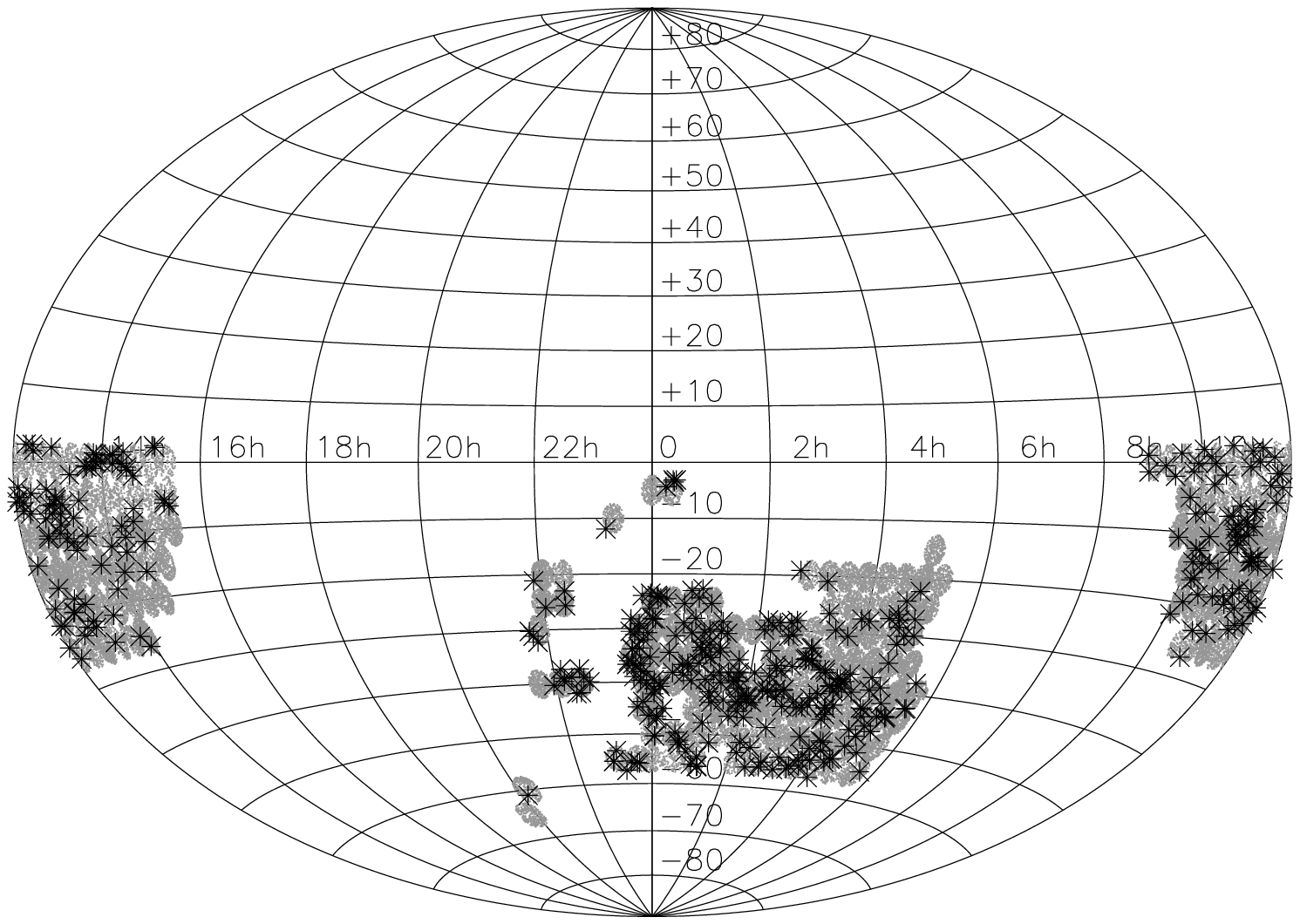}{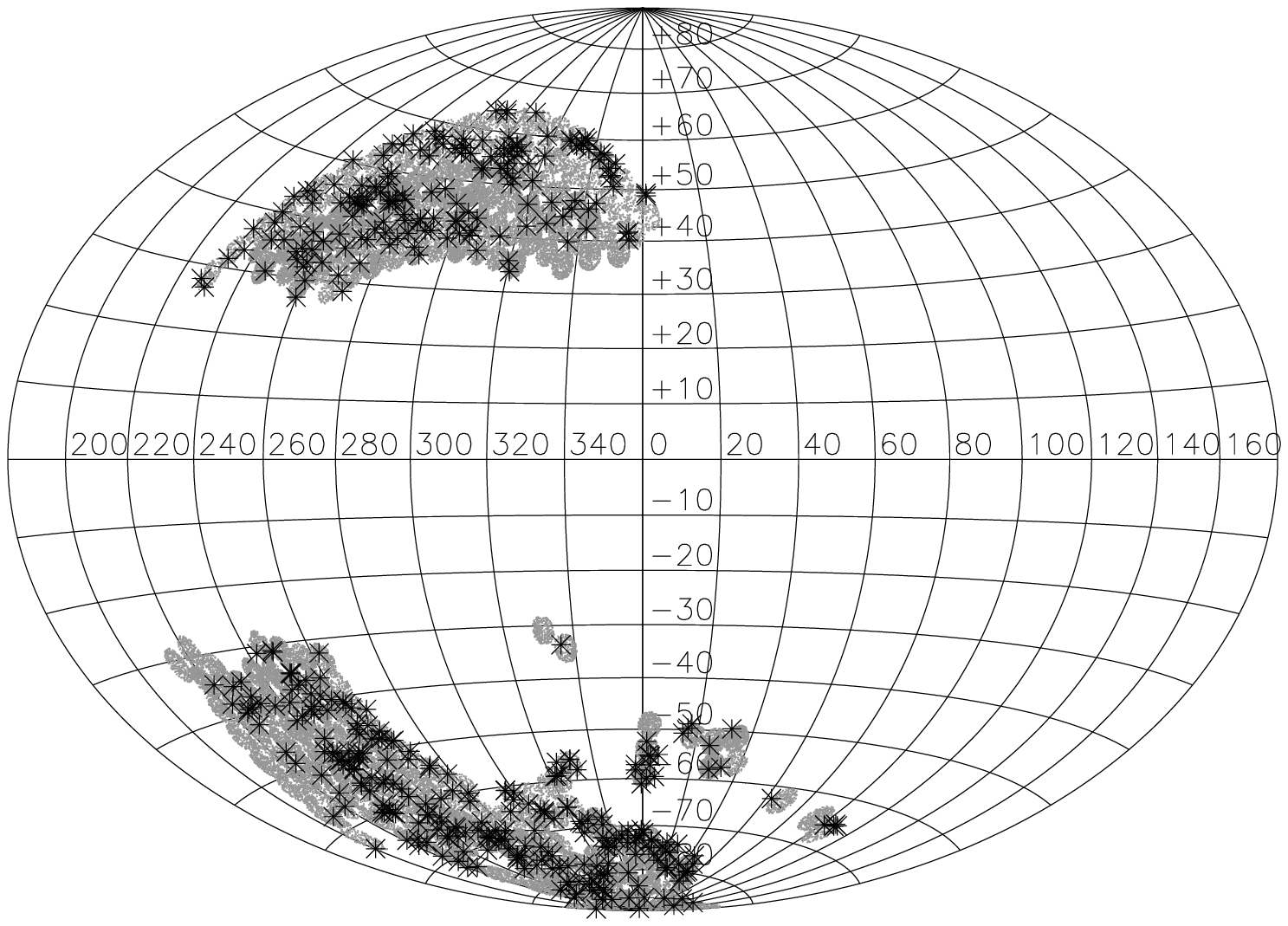}
\caption{Sky positions of APM galaxies (grey points) and 
Hamburg-ESO QSOs (black asterisks)
used in this study.  {\it Left:} Equatorial coordinates.  
{\it Right:} Galactic coordinates.}  
\label{sky}

\end{figure}

Because the catalogue is derived from a photographic Schmidt survey,
vignetting is a factor on individual UKST plate fields.  It reduces
the plate density of objects, but the magnitude of the effect is 
negligible within a radius of $2\fdg7$ from respective UKST
field centers.  Hence in order to avoid the contamination of the galaxy-QSO
cross-correlation, we remove all galaxies lying more than $2\fdg7$ 
from their respective field centers from our study.  The locations of the
galaxies used in our study, along with QSO positions can be found in
Figure~\ref{sky}.

\begin{figure}
\epsscale{.8}
\plotone{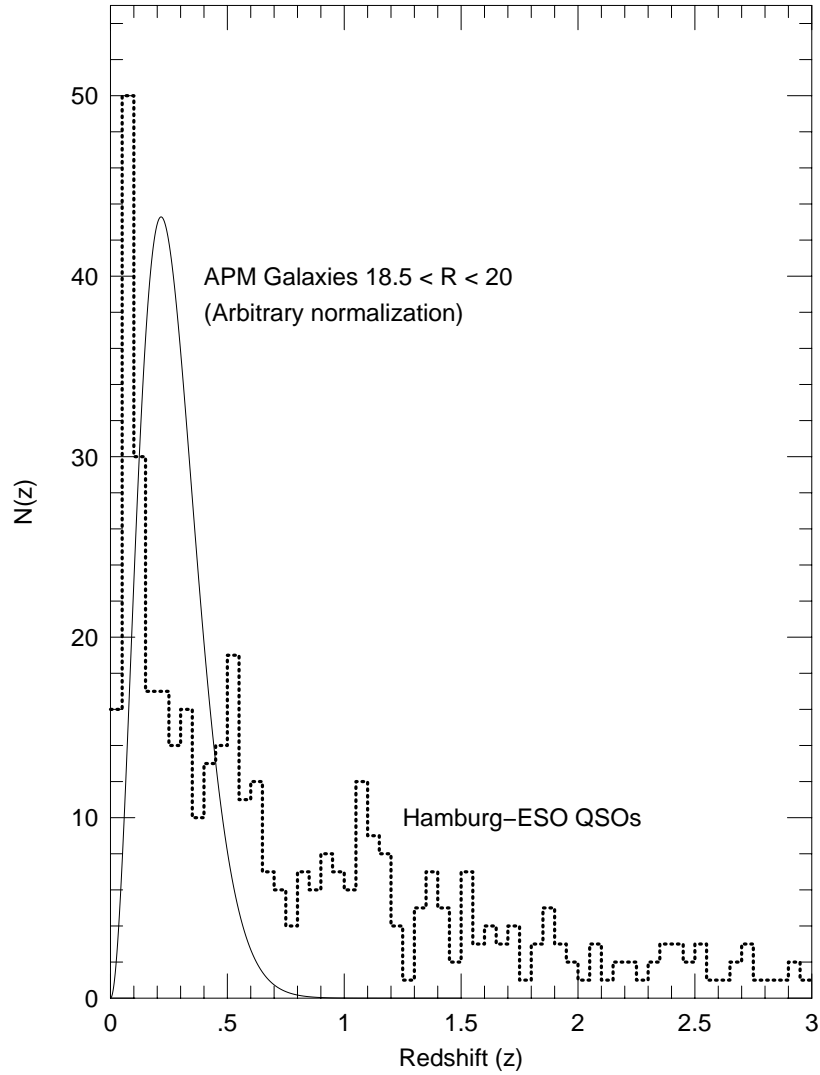}
\epsscale{1}
\caption{Redshift distributions of Galaxies and QSOs used in this study.  The
histogram represents the QSOs in the Hamburg-ESO QSO sample while the solid
line shows the estimated redshift distribution (with arbitrary normalization)
for our sample of roughly $4.5 \cdot 10^{6}$
APM Galaxies with magnitudes $18.5 \leq R \leq 20$, 
following \citet{MES96}.}  
\label{gqz}

\end{figure}

We estimated the redshift distribution of galaxies in our magnitude 
cut using the fit given by \citet{BE93}. The parameters of this fit match
the observed magnitude-binned galaxy redshift distributions from the Stromlo/APM
redshift survey \citep{LPEM92a, LEPM92b}, as well as the Durham/Anglo-Australian
Telescope faint galaxy redshift survey conducted by
\citet{BES88} and the LDSS faint galaxy survey by \citet{CETH90, C93}. 
(Please see Section 4.1 of \citet{MES96} for more details regarding the
development of this fit.)  The resulting distribution
(with arbitrary normalization), shown as a solid curve in Figure~\ref{gqz}, 
peaks at a redshift of $z\approx 0.25$ and falls off almost entirely 
by $z \sim 0.5 - 0.6$.  The galaxy sample does overlap the low-redshift portion
of our QSO sample (see Section~\ref{heso}), 
which we retain to investigate differences between the
cross-correlation signals due to physical associations and magnification bias.

\subsection{Hamburg-ESO Survey for Bright QSOs}
\label{heso}
 
The Hamburg-ESO Survey for Bright QSOs is a survey that employs 
slitless spectroscopy on the ESO 1m Schmidt Telescope to find 
bright QSOs over an area that will eventually cover roughly $5000~\sq\degr$ of 
the southern sky. See \citet{w96} and \citet{RKW96} for details regarding 
the methods and techniques used in the development of this survey.  
Currently, a preliminary portion (roughly half) of the final dataset is 
available to the public. This subset contains 415 bright QSOs and 
Seyfert I nuclei with magnitudes in the 
range $13 \lesssim B_{J} \lesssim 17.5$ and redshifts $z < 3.2$  that span 
an effective area of roughly $3700~\sq\degr$ \citep{w00}.  Crowding of sources, 
especially on fields of low Galactic latitude, does result in the loss of
spectra for roughly $\sim 30\%
$ of the spectra on the fields.  Nevertheless, the Hamburg-ESO 
survey QSO sample is particularly useful for this work because it is $99 \% 
$ spectroscopically 
complete over the well-defined ranges of magnitudes and redshifts given 
above and because it is devoid of redshift- and magnitude-related selection 
biases.  The redshift distribution of Hamburg-ESO QSOs is shown as a dashed
line histogram in Figure~\ref{gqz}.

Though Hamburg-ESO QSOs were identified on plates containing slitless spectra
corresponding to ESO/SRC
field position (not the UKST), QSOs selected for this study were those 
whose positions lie within the $2\fdg7$
non-vignetted regions of the nearest respective UKST fields from which we had 
garnered our galaxy data.  There is more overlap between UKST fields at high
declinations than there is near the Celestial Equator.  Therefore some
high-declination QSOs lie within $2\fdg7$ degrees of multiple field centers.  In
these cases, we associated the QSO with the plate corresponding to the nearest
field center.   Those QSOs that lay further than $2\fdg7$ from any field center
were not included in this study.

The QSOs that fall within the apparent magnitude range found in the Hamburg-ESO 
Catalogue are typically intrinsically brighter than $M^{*}$ at most (especially
at higher) redshifts.
This implies that the catalogue samples the steep portion of 
the QSO number-magnitude count distribution and that one would expect to 
measure positive galaxy-QSO cross-correlations 
{\it \`{a} la} Equation~\ref{lens}.

\section{Analysis}
\label{proc}

\subsection{Estimation of the Correlation Function}
\label{est}

The estimation of the galaxy-QSO cross-correlation is in many regards very similar
to the determination of the galaxy auto-correlation, which has been conducted by
many studies in the past \citep{P80, H82, MES96} and refinements have been
incorporated in the technique \citep{H93, LS93}, but there are some notable
differences as well.  

Unlike autocorrelation estimates, the
galaxy-QSO cross-correlation involves two separate catalogs, each with its own
selection and window functions.    In many cases, 
the selection function could also be an issue because one does not 
know {\it a priori} either the 
underlying galaxy or quasar distribution from which survey samples are 
obtained.  Furthermore, the window function of a survey could also influence 
the correlation estimates by introducing geometrical effects that conspire 
with the sky densities of objects to skew correlation estimates relative to the
values one would obtain from fair samples that very well
characterize the true distributions of galaxies and QSOs. 

An additional complication is the fact that the number of galaxies in our
magnitude cuts from the APM catalogue is on the order of $\sim 10^{4}$ times 
greater than the number of QSOs in the Hamburg-ESO QSO catalogue.  
As a result, the uncertainty due to Poisson noise is greater relative to the
signal for galaxy-QSO cross-correlations
than it is for galaxy-galaxy auto-correlations.

A number of autocorrelation estimators can be found in the literature,
including:
\begin{equation}\omega_{GG1}(\theta) = \frac{DD}{\langle DR \rangle} - 1,\end{equation} 
\begin{equation} \omega_{GG2}(\theta) = \frac{DD \cdot \langle RR \rangle}{\langle DR \rangle^{2}} 
 - 1,\end{equation} 
 \begin{equation} \omega_{GG3}(\theta) = \frac{DD - 2\langle DR \rangle + \langle
RR\rangle}{\langle RR\rangle },\end{equation} and
\begin{equation} \omega_{GG4}(\theta) = \frac{DD}{\langle RR \rangle} - 1.\end{equation}
Here, $D$ and $R$ refer to whether the galaxies were drawn from 
direct or randomly generated samples, respectively, in order to generate
the pair counts $DD$, $DR$ and $RR$.  Angled brackets 
represent averages over a large number of realizations (100 realizations 
in the case of this work) of the total pair counts between 
combinations of random (artificial) galaxies.  Within each realization,
we use a number of randomly generated objects as there are real objects in order
to preserve normalization (this is true in the case of cross-correlations as
well).

The most common estimator, $\omega_{GG1}$, has largely replaced the 
original one, $\omega_{GG4}$, in the literature and 
is often used in both auto- and cross-correlation (in a variant form) studies. 
This is because $\omega_{GG4}$ is most severely affected by
window effects of the correlators mentioned here.  It is a fact that has been 
known for quite some time, and has resulted in the development of the newer 
correlation estimators, $\omega_{GG2}$ and 
$\omega_{GG3}$, that \citet{H93} and \citet{LS93} developed in order to 
minimize the influence of contamination by the window and selection functions.
These latter functions have the useful property that they are dependent on the
galaxy-catalogue correlation function to higher orders than $\omega_{GG1}$ and
should be less affected by selection and window effects as a result.  

We have adapted these autocorrelation estimators for use in our 
cross-correlation study:
\begin{equation}\omega_{1}(\theta) = \frac{D_{G}D_{Q}}{\langle D_{G}R_{Q}
\rangle} - 1,\end{equation} 
\begin{equation} \omega_{2}(\theta) = \frac{D_{G}D_{Q} \langle R_{G}R_{Q}
\rangle}
{\langle D_{G}R_{Q} \rangle \langle R_{G}D_{Q}\rangle} - 1, \end{equation}
\begin{equation} \omega_{3}(\theta) = \frac{D_{G}D_{Q} - \langle R_{G}D_{Q}\rangle 
- \langle D_{G}R_{Q} \rangle +
\langle R_{G}R_{Q}\rangle }{\langle R_{G}R_{Q}\rangle },\end{equation}  
and,
\begin{equation} \omega_{4}(\theta) = \frac{D_{G}D_{Q}}{\langle R_{G}R_{Q}
\rangle} - 1, \end{equation} 
where the subscripts $G$ and $Q$ indicate whether the direct or random samples
represent galaxies or QSOs, respectively.  

Like the galaxy autocorrelation estimators, each of the cross-correlation
estimators are dependent to varying degrees on catalogue window functions and
selection effects.  Therefore, we have extended the formalism describing these
dependencies found in \citet{H93} to the 2D auto- and cross-correlation 
estimators. Following the \citet{H93} formalism for a 2D catalogue, let us first
define the catalogue selection function, $\Phi_{x},$ which could be a function of
object brightness, color, position, etc., for catalogue, $x$, which would be
either $x = G$ or $x = Q$ for galaxies or QSOs, respectively.  
Unlike \citet{H93}, we will assume that
the positional dependence of $\Phi$ is 2-dimensional.  If $\nu$ is the surface
density of objects within a catalogue, then the total number of objects within
the catalogue, $x,$ can be determined with \begin{equation} N_{obs,x} =
\Phi_{x} \nu_{x} ,\end{equation} so that if a catalogue contained all of the objects
that lay within the bounds of the survey region, the selection function would be
$\Phi = 1.$  $\Phi$ is technically a discrete function, but it is entirely
possible to treat it as a continuous function, 
especially with large numbers of catalogue objects. 

If the catalogue window function is $W_{x},$ then $w_{x}$ is defined
so that \begin{equation} W_{x} \equiv w_{x}\Phi_{x}, \end{equation} and the
cross-pair
window is \begin{equation} W_{GQ}  = w_{GQ}\Phi_{G}\Phi_{Q}. \end{equation} 
The pair weighting, $w_{GQ},$ need not be related to the point
weightings, $w_{G}$ and $w_{Q}$.  Then the direct pair counts will involve $N_{obs,x},$ weighted by a function,
$w_{x},$ and the direct-direct pair counts for
cross-correlation analysis will be given
by \begin{equation} D_{G}D_{Q} = w_{GQ}N_{obs,G}N_{obs,Q} = 
w_{GQ}\Phi_{G}\Phi_{Q}\nu_{G}\nu_{Q}, \end{equation}  
while direct-random pairs will be given by either \begin{equation}
D_{G}R_{Q} = w_{GQ}N_{obs,G}\Phi_{Q} = w_{GQ}\Phi_{G}\Phi_{Q}\nu_{G}
\end{equation} or by \begin{equation}R_{G}D_{Q} = w_{GQ}N_{obs,Q}\Phi_{G} =
w_{GQ}\Phi_{Q}\Phi_{G}\nu_{Q} ,\end{equation} and finally the random-random
pairs are given by \begin{equation} R_{G}R_{Q} = W_{GQ} =
w_{GQ}\Phi_{G}\Phi_{Q}.  \end{equation}

If $\bar{\nu}_{G}$ and 
$\bar{\nu}_{Q}$
are the mean sky surface densities of galaxies and QSOs lying within respective
catalogue windows, then the degree to which the represent the true full-sky
average surface density, $\bar{\nu},$ can be estimated with the sample 
overdensities $\sigma_{G}$ and $\sigma_{Q}$, where \begin{equation} 
\bar{\sigma}_{x} = \frac{\bar{\nu}_{x} - \bar{\nu}}{\bar{\nu}} \end{equation}
for catalogue $x$.  Fair samples that very well represent the true sky density
of objects will have sample overdensities $\bar{\sigma}_{x} \approx 0.$ 
Similarly, the object-catalogue correlation, $\phi_{x},$ which is defined by
either \begin{equation} \phi_{G} \equiv \frac{\langle W_{GQ}\sigma_{G} \rangle}
{\langle W_{GQ}\rangle }, \end{equation} or \begin{equation}\phi_{Q} \equiv 
\frac{\langle W_{GQ}\sigma_{Q} \rangle}{\langle W_{GQ} \rangle},\end{equation}
should also be zero for a perfectly fair sample because the true object 
surface density in a sample, $\nu_{x},$ would be equal to the true cosmic mean,
$\bar{\nu}$.  

Considering that the measured cross-correlation function will be, in general, a
biased estimate of the true cross-correlation, we can define the windowed
correlation, \begin{equation}\hat{\omega}_{GQ} \equiv \frac{\langle
W_{GQ}\sigma_{G}\sigma_{Q} \rangle}{\langle W_{GQ} \rangle}, \end{equation}
which is equivalent to an unbiased cross-correlation in ideal circumstances.

The results, giving the correlators and their influence by 
the catalogue sample overdensities, $\bar{\sigma}_{G}, \bar{\sigma}_{Q}$, 
the catalogue-window
correlations, $\phi_{G}, \phi_{Q},$ and the object-catalogue correlations,
$\hat{\omega}_{GQ}$, are shown in Table~\ref{estcat2}.  Additional estimators
($\omega_{5} \dots \omega_{7}$) are also included.  While these three estimators
are not useful
for determining the galaxy-QSO cross-correlations, they are helpful in gauging
the degree to which randomly generated object distributions match real object
distributions for modeling purposes.  The most problematic terms  in the third
column of Table~\ref{estcat2} are the first-order terms 
involving the catalogue-window correlations (the $\phi$'s).  
It is then apparent that just like
$\omega_{GG4}$, the cross-correlation estimator, $\omega_{4}$, is inferior
because of its greater dependence on the catalogue-window correlations, whereas
$\omega_{3}$ and $\omega_{3}$ are only dependent on them to second order.

Direct pair counts were determined by counting all of the galaxies within angular
separation bins $\theta \pm \frac{\delta\theta}{2}$ around QSOs,
 using galaxies that were taken from one field per QSO only. In addition to
 eliminating the effect of vignetting, this eliminates the
 need to account for plate to plate photometric sensitivity variations across
 the APM catalogue.
Furthermore, all pairs were drawn from the inner $2\fdg7$ portion of each 
UKST field.  Though it does limit the angular scale over which we can estimate 
the cross-correlation to scales $\theta \leq 5\fdg4$, this scale is rather large
when compared to the scales over which prior studies have investigated and it
prevents the 
contamination of the correlation signal by the plate to plate variations in 
the APM Catalogue photometric calibration and it serves to mitigate the
influence of vignetting on the plate-density of galaxies.  

The pair counts involving randomly generated QSOs 
and galaxies were also treated in a manner that limited the inclusion of 
geometric effects.   For each QSO in our sample, 100 {\it Circularly-Distributed
Random} (CDR) QSOs were generated
that lay in a circle about the field center with a radius equal to the
separation between the real QSO and the field center, and pair counts were
averaged over the 100 realizations.  Using CDR QSOs
 ensures that radial sensitivity gradients that may be present on the 
 APM Schmidt plates cancel out, as will 
much of the edge effect introduced by the $2\fdg7$ radius boundary. Random
galaxies for use in determining the $\langle R_{G}R_{Q} \rangle$ and 
$\langle R_{G}D_{Q} \rangle $ terms that are found in the estimators
$\omega_{2}, \omega_{3}$ and $\omega_{4}$ were created by randomly
scattering points on the field so that the probability of a random galaxy
being generated at some spot on the field as equal for all points within
$2\fdg7$ of the field center and 0 outside that radius.  The number of
randomly generated galaxies and the number of real galaxies on the 
corresponding
UKST field were the same in every realization in order to preserve normalization.

\subsection{Angular Cross-Correlations}
\label{c}

With combinations of pair counts in hand, we estimated the angular 
cross-correlations using each of the four estimators from the previous section
with subsets of QSOs from a number of redshift bins of width $\Delta z = 0.2$ 
in the range $0 \leq z \leq 2$.   
This is the first time that all four of these cross-correlation estimators have 
been compared using the same QSO and galaxy data sets and each  estimator 
gives very similar results in each subset.  A typical comparison of each of the
estimators and the magnitude of their respective RMS uncertainties is shown in 
Figure~\ref{corr} using QSOs within a redshift range of 0.8 - 1. 
As expected, the greatest discrepancies between them are at the smallest and 
largest galaxy-QSO separations where the variance (also shown) is highest. 
At small separations, the variance is due largely to Poisson fluctuations in 
the small number of galaxies located very near QSOs while the rise in variance 
at large pair separations is due both to increasing relative Poisson 
uncertainties from the increasing importance of edge effects near the 
limiting separation scale of $5\fdg4 = 324\arcmin$ of our study   use the
dispersion among the individual QSO estimates as a conservative 
measure of uncertainty rather than the error in the mean due to the 
possible presence of the various systematic effects that were mentioned in the
previous section.

Among the individual correlators, $\omega_{4}(\theta)$, deviates the most from 
the others, especially at small and large separation scales where the stronger 
dependence of $\omega_4$ on the window function becomes important. This is not
surprising as $\omega_4$ has long been considered an inferior estimator due to
its relatively strong dependence on selection effects and window functions
\citep{H93}.  On the other hand, $\omega_1$, $\omega_2$, and $\omega_3$ all 
yield very similar results across, which means that $\omega_1$ is apparently still an 
acceptable choice.  To be sure, \citet{LS93} and \citet{H93} have shown 
that $\omega_2$ and $\omega_3$ are less dependent on window and selection 
effects than $\omega_1$, but they also require significantly more computational
time in order to sample the additional random galaxy population that $\omega_1$
does not require. Hence if the cost of computation time exceeds the benefit of
marginal signal improvement, then perhaps $\omega_1$ would prove useful.

Furthermore, that each of the estimators, with their different dependences on 
the catalogue-object cross-correlations and relative over- or underdensity
relative to the true object populations, are so similar implies that to a
good degree both the APM galaxy catalogue and the Hamburg-ESO QSO catalogue are
fair samples of their respective true object populations.  This is significant
because a number of earlier studies (see Table 1) that have used the APM 
galaxy catalogue have found significantly stronger cross-correlations than 
expected from models.  If in some way, the APM did not represent a fair 
sample, statistical associations could conceivably result.  

Finally, the angular cross-correlation measurements are quite significant.
While it is true that in the case of a single QSO on an APM field, 
the galaxy counts in adjacent annular
separation bins will be correlated at some level because of the clustering of
galaxies on the sky and because of sensitivity variations on the plate, this
effect is likely small because we have seen that each of the different
estimators yield similar results despite varying sensitivity to the underlying
galaxy distribution. And while individual separation bins are only detected 
to $~ 2\sigma$ on
Figure~\ref{corr}, given that there are 415 QSOs in the Hamburg-ESO sample,
which covers $3700~\sq\degr ,$ the average separation between QSOs is
$\sim 3\degr $, so each of the small-separation angle bins should be nearly
statistically independent because there should be little overlap of
small-separation annuli around QSOs.  As a result, considering that the small
separation bins all show positive correlations and that they also follow an
overall trend inconsistent with zero slope indicate that the detection of 
positive galaxy-QSO cross correlations is rather strong. 

\subsection{Cross-correlations vs. Source QSO Redshift}
\label{ccz}

If the cross-correlations discussed in Section~\ref{c} are due to magnification
bias, then their amplitudes should monotonically rise with increasing 
redshift beyond
the lensing plane (at $z \approx 0.25$) \citep{DB97,BS01, MB02}.  
We integrated the angular 
cross-correlations over scales up to $30$ and $60\arcmin$ in order to 
increase the signal to noise ratio enough so
that we could divide our QSO sample into a number of subsets based on QSO
redshift and explore the $z-$dependence of the associations.  

Figure~\ref{xvz} shows the integrated cross-correlation vs. redshift.  
Because of low QSO numbers per bin, we have used very wide $(\Delta z = 0.4)$ 
redshift bins and we have therefore oversampled to show the degree of 
variation with redshift.  We do detect a positive signal on the order of
$1 - 3\%
$ over all redshifts beyond the lens redshift, 
which is qualitatively consistent with magnification bias, and as expected, the
Poisson noise is greater for the $30\arcmin$ plots than for
the $\theta < 60\arcmin$ plot. This can be complicated by the integral
constraint, \begin{equation}\frac{1}{\pi\theta^{2}}\cdot
\int_{0}^{5\fdg4}{\omega(\theta) \cdot 2\pi\theta d\theta} = 0. \end{equation}
Positive correlations at small separations must be balanced by
negative correlations at higher separations.  

Since the correlators were found to be very similar in the previous section,
one would expect the same of the integrated estimators in Figure~\ref{xvz}. 
This is generally the case, although the relative differences between the
integrated estimators is a bit greater than the relative differences between the
estimators themselves.  Because \begin{equation}\omega(< \theta_o) =
\frac{1}{\pi\theta_{o}^{2}}
\int_{0}^{\theta_{o}} \omega(\theta) \cdot 2\pi\theta d\theta ,
\end{equation} most of the signal contribution to $\omega(< \theta_o)$ comes from
angular scales very near $\theta_o$.  Perhaps the differences between the
integrated correlation function is then due to the modification of the angular
correlation profile by the integral constraint.  Minor differences between
angular correlations on scales $\approx \theta_o$ would be
exacerbated by the geometric scaling $\sim 2\pi\theta_o.$
 
The overlap between our galaxy and QSO samples at low redshift (Figure~\ref{gqz})
means that the low redshift ($z \lesssim 0.4$) signal in Figure~\ref{xvz} is due
to physical associations rather than magnification bias.  The $\theta < 
30\arcmin$ plot does in fact show a drop in the cross-correlation signal over
redshifts in this range, along with a rise at redshifts greater than 0.4.  This
seems to indicate a fall off in the signal due to physical associations between
galaxies and QSOs in front of the lensing plane, followed by an increase in the
magnification bias signal involving QSOs beyond the lensing plane.  The $\theta
< 60\arcmin$ plot does not show a significant dip between the physical and lensing
associations.  Perhaps this is due to a combination of the oversampling in
redshift and due to the integral constraint.  Separations of $60\arcmin$ are of the
same order of magnitude as the $5\fdg4$ limiting scale of our study and the
integral constraint may be altering the correlations on these scales.

\subsection{Confidence Tests}
\label{ct}

Individual angular cross-correlation functions from redshift bins with width
$\Delta z = 0.4$ tend to be a bit noisy.  Therefore, it is imperative to
create a significance test for the integrated correlation function.  
Working toward this end, we have developed a Monte Carlo algorithm that
generates a large number of realizations of the integrated 
cross-correlation measurements in the same manner as those from the last
section.  It utilizes our APM galaxy sample along with a large number of 
artificially generated QSO catalogues in order to determine the significance 
of the integrated correlation funtion measurements from the last
section.  

The algorithm works as follows:
First, for each QSO in the Hamburg-ESO Catalogue, we generated one random QSO 
so that the probability lying at a particular point within $2\fdg7$ of the field center
for this first type of QSO was equal for all points and 0
outside of that radius (as we did for our randomly generated galaxies, see
Section~\ref{est}).  
We shall call these {\it Uniformily-Distributed Random
} (UDR) QSOs.  Each UDR point was located on the same field as the real QSO that
it replaced.  Furthermore, in order to preserve the number of QSOs per redshift
and magnitude bin, 
each was also attributed the same magnitude and redshift as the replaced real
QSO.  Next, for each UDR QSO, we generated 100 CDR QSOs (see Section~\ref{est})
whose positions were
distributed along a circle about the field center with a radius equal to
the distance between the UDR QSO and the field center, as in our real-QSO
analysis. Finally, we determined the correlation function,$\omega_1$, 
from the resulting pair counts.  This process was
repeated 1000 times, effectively creating 1000 random realizations of the
out cross-correlations.  We interpret the distribution in values of the
artificial
cross-correlation functions to be indicative of the magnitude of the uncertainties
in our measurement of the cross-correlation function.  

Figure~\ref{xvz} shows
the $50, 75,$ and $99\%
$ confidence limits that we derive from this interpretation, corresponding to
the ranking of the values of 1000 cross-correlation realizations in each
redshift bin.  As the diagram indicates, the physical association signal is
detected to high significance, especially in the lowest redshift bin,
corresponding to source QSOs with redshifts $\leq 0.4$. The magnification bias
signal seems to be most significantly detected in  the $\theta < 30\arcmin$
frame at redshifts around 0.8 and 1.3, although the overall detection is
somewhat marginal - with confidence levels on the order of $75\%
.$  

We do consider this Monte Carlo confidence analysis to be rather
conservative, however.  This is due to the fact that randomly generated QSOs can
be either positively or negatively associated with APM galaxies and the
confidence contours illustrate the ranking of individual realizations.  The
$99\%
$ contour therefore samples the very extremes of the underlying distribution. 
which seems to have a rather wide tail. 
Hence we believe the confidence limits are somewhat {\it underestimated}.

It is also useful to compare the galaxy-QSO results with those of
cross-correlations between objects for which there should be no net
cross-correlation.  Therefore, we selected objects listed as stars in the APM
catalogue with magnitudes $16 \leq R \leq 17.5$.  Aside from the contamination
of our stellar sample by 
galaxies on the faint side of our magnitude and object class cut, there is no
reason that Galactic stars should be correlated in any way with QSOs that lie at
cosmological distances.  Figure~\ref{xvzs} shows the strength of the integrated
Star-QSO cross-correlation vs. redshift that was obtained in the same manner
as the galaxy-QSO cross-correlation.  No real dependence on redshift can be seen
as each of the cross-correlation estimators oscillate above and below $0$ in a
manner that would be expected for random noise.  Indeed, the star-QSO 
cross-correlation can serve to provide an independent indication of the 
noise level for the galaxy-QSO signal that is consistent with our Monte 
Carlo confidence estimate.  

\subsection{Modeling}
\label{model}

We modeled the signal in Figure~\ref{xvz} based on a linear CDM large scale structure 
lensing model, given by \citet{BS01}, which gives the expected correlation function
from an assumed mass power spectrum.  This particular model is a simplified
version of that given by \citet{DB97} in which an approximation is made by
assuming that all galaxies are found at their median redshift.  This
approximation incurs errors on the order of $10\% 
$.  The mass power spectrum used for our model was Case 1 obtained from the
Appendix G of \citet{BBKS} (BBKS): A CDM power spectrum with adiabatic fluctuations.  
We assumed a value of the present-day mass fluctuation normalization of $\sigma_{8} = 1$, 
a matter density 
parameter of $\Omega_m = 0.3$, cosmological constant 
$\Omega_{\Lambda} = 0.7$, and expansion rate $h = 0.75$,
resulting in a shape parameter of our
modeled mass power spectrum was $\Gamma = 0.225$.  The power spectrum at the
lens plane redshift was then determined by scaling the present-day
power spectrum with a factor of $(1 + z_{lens})^{-2}$.  

Using our mass power spectrum and the assumption that all of the galaxies in our
sample lie on a plane with a redshift $z = 0.25$, we determined the expected
strength of the angular cross-correlations using the \citet{BS01} approximation
while varying the source redshift over the
range represented by our QSO sample.  Each of the correlation signals were
integrated over separations up to $30$ and $60\arcmin$, respectively. 
The model predicts an increase in the strength of the signal with
source distance, provided that the sources lie behind the lensing plane. This
can be seen in Figure~\ref{xvz}, which agrees well qualitatively with 
the measured signal from the combination of Hamburg-ESO QSOs
and APM galaxies.  However, consistent with a large number of other studies
(see the Introduction), the integrated modeled correlations were 
significantly lower (roughly a factor of 20) than our measured correlations. We
will explore this further in the Discussion section.
  
\subsection{Correlation scales}
\label{cl}  

As we have mentioned earlier, we see in  Figure~\ref{xvz} that QSOs with 
redshifts $z \lesssim 0.4$ are physically
associated with the APM galaxy sample, while those at higher redshift are likely
affected by magnification bias.   We have separated our QSO sample into two
redshift ranges in Figure~\ref{cz}, $z \leq 0.2$ and $0.6 \leq z \leq 1.0$, in
order to see whether the differing assumed physical mechanisms result in
different correlation functions.  Outside of differences on angular
scales greater than $130\arcmin$, the low and high redshift sample have very
similar profiles.  Both possess `correlation lengths' 
(defined here as $\theta_o$ such that $\omega(\theta_o) = 0$, as opposed to the
standard definition of the physical length, $r_o$, where $\xi(r_o) = 1$) 
of $\sim 110\arcmin$, which corresponds to scales of roughly 23 Mpc at the 
redshift of our APM galaxies ($z \approx 0.25$) 
(assuming $h = 0.75$).  These scales are much
larger than the typical sizes of groups and clusters and they seem to
indicate that both physical and lensing associations are due to 
Large Scale Structure.  That the physical and lensing associations have roughly the same correlation
length may be due to the fact that the same structure is likely responsible for
both signals.  

\subsection{Galactic Dust}
\label{gd}

With such a large disagreement between the positive cross-correlations that are
observed in this study with model expectations, it is appropriate to investigate
whether these cross-correlations could in fact be generated by Galactic dust.
If the QSOs in the Hamburg-ESO sample happen to lie in portions of the sky that
are relatively unobscured by intervening Galactic dust, it could conceivably
explain the amplitude discrepancy. \citet{RWH94} showed that the
overdensity due to differences in levels of extinction, $\delta A_{V}(\theta),$
between two lines
of sight separated by an angle, $\theta,$ is given by
\begin{equation} q(\theta) = 10^{-2.5 \cdot \alpha_{m} \delta A(\theta)} \approx 1 - 5.76 
\cdot \alpha_{m} \delta A(\theta). 
\label{dusteq} \end{equation}  If dust were the sole cause of the observed
cross-correlations, then the maximal overdensity produced by dust would be
roughly proportional to the differential extinction.

The \citet{RWH94} analysis determined the extremes in extinction across LBQS
fields using the \citet{BH82} maps and found that the observed galaxy-QSO
cross-correlations between LBQS QSOs and APM galaxies were in fact 
anti-correlated with the cross-correlations that would have been expected 
if they were due solely to Galactic extinction.  

We used a slightly different technique in which we interrogated
the \citet{SFD98} Galactic $E(B - V)$ reddening maps using the IDL {\tt dust\_getval}
routine, which is provided by \citet{SF99},  
to derive and compare the differences in
extinction levels between Hamburg-ESO QSO positions and large-scale averages
within surrounding annuli with inner and outer radii of $135\arcmin$ and
$150\arcmin ,$ respectively. We define the 
differential extinction
by \begin{equation} \delta A_{V}(\theta) = A_{V}(<15\arcmin) - A_{V}(135 - 150\arcmin).
 \end{equation} 
The \citet{SFD98} reddening map has a resolution of $6\fm 1 \times 6\fm 1$ and local reddening
values were interpolated from the adjacent surrounding pixels.
We determined extinction levels from the \citet{SFD98} reddening map
by adopting a single value for the ratio of total to 
selective extinction $R_{V} = 3.1$ for every Hamburg-ESO QSO.

\begin{figure}
\epsscale{0.8}
\plotone{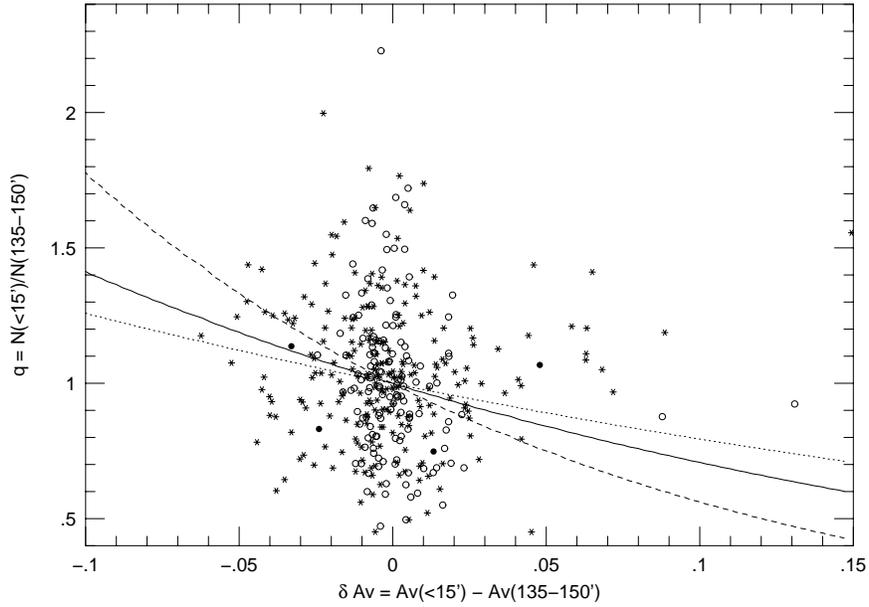}
\epsscale{1.0}
\caption{The observed overdensities of galaxies within $15\arcmin$ of QSOs
with respect to galaxies between $135 - 150\arcmin$ is plotted here versus the
difference in Galactic extinction between the same separation scales.  The
points are divided between high- (60 - 90\degr), mid- (30 - 60\degr), and 
low-Galactic Latitude (0 - 30\degr) samples (empty circles, stars, and filled
circles, respectively).  The points appear completely uncorrelated.  In addition
the expected maximal overdensities are plotted as a function of differential
extinction for cases in which the QSO cumulative number-magnitude count 
slope $\alpha_{m}= $ 0.6 (solid), 0.4 (dotted), and 1.0 (dashed). (See
Equation~\ref{dusteq}.)
}
\label{gdust}
\end{figure}

Meanwhile, we determined the galaxy number densities from the same regions and
measured the relative overdensities between the QSO positions and the field:
\begin{equation} q = \frac{N(<15\arcmin)}{N(135 - 150\arcmin)}. \end{equation}
The reason for the specific choice of the annular region size was that it
represents the approximate angular scale at which the galaxy-QSO
cross-correlations, $q(\theta) - 1 = \omega(\theta)$ are near zero for our
sample.  Figure~\ref{gdust} shows that the relative overdensities and the
differential extinctions between the QSO positions and the field are
uncorrelated.  In addition, the solid, dotted and dashed lines in 
Figure~\ref{gdust} show the maximal overdensities that would be expected from
dust extinction alone, given slopes of $\alpha_{m} = 0.6, 0.4,$ and $1.0,$
respectively.  The slope of 0.6 corresponds roughly to the slope of the QSOs in
our sample, while 0.4 corresponds to the critical slope at which there would be
no expected lensing signal.  If dust were the sole cause of the
cross-correlation signal, one would expect that the magnitude of the
overdensities (underdensities) would lie somewhere between a value of $q = 1$
and the curve derived using the appropriate QSO sample slope.  
Yet, the most of the QSO overdensities fall outside this bound.

\section{Discussion and Conclusions}
\label{disc}

\begin{figure}
\epsscale{0.8}
\plotone{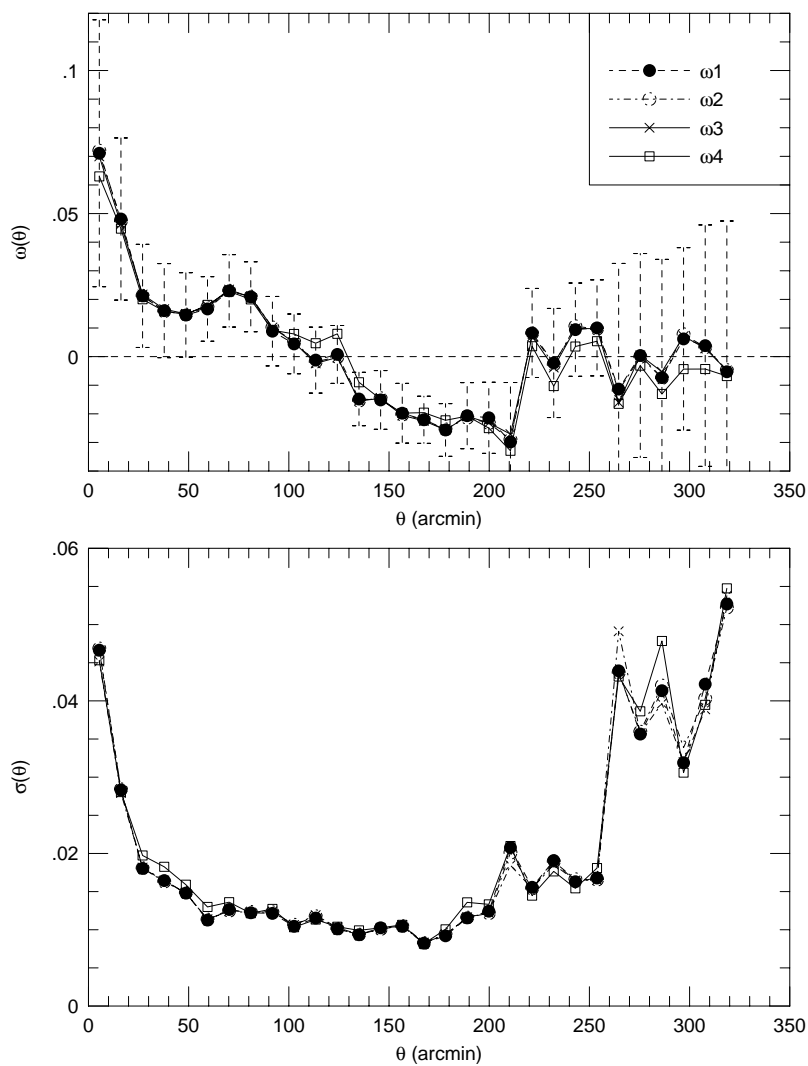}
\epsscale{1.0}
\caption{A comparison between correlation function estimators, using a sample
of QSOs with redshifts $0.8 < z < 1.0$ and magnitudes $B_J \leq 17.5$.  
{\it Top:} Angular correlation function
vs. pair separation angle.  The correlators $\omega_1, \omega_2$ and $\omega_3$
are very similar in this case.  {\it Bottom:} RMS Uncertainty in the correlation function
calculated using the distribution of correlation functions of individual QSOs within
the sample.  The errorbars in the top panel show the RMS uncertainty for
$\omega_1$.}
\label{corr}
\end{figure}

An important result of this work is the detection of a net positive correlation
signal over the entire redshift range for which we have significant numbers of
QSOs (see Figures~\ref{corr} and \ref{cz}).  Though our Monte Carlo 
simulations suggest a modest level of confidence for the integrated 
correlations, our having detected a consistently positive signal over all 
redshift  intervals suggests that perhaps a 
somewhat higher level of confidence could be placed on the result.  The Monte
Carlo results also belie the fact that the {\it angular} 
cross-correlations from various redshift intervals are quite significant 
and that they indicate strong detections
of positive galaxy-QSO cross-correlations.  We
anticipate that when the final Hamburg-ESO QSO Catalogue is finished,
effectively doubling the size of the current sample, the total S/N ratio 
should in principle improve by a factor of roughly $\sqrt{2}$, which would 
lend a significantly higher level confidence in the cross-correlation
measurements.  

The measurements of statistical associations 
between APM galaxies and Hamburg-ESO QSOs occur 
on angular scales $\theta \lesssim
110\arcmin$.  The overall shape of the correlations for physical and lensed
associations are very similar.  Meanwhile, the correlation length (equivalent for both types of
associations, corresponding to $\sim 20$Mpc
scales at $z_{lens} \approx 0.25$  suggests that large scale structure 
is doing the lensing rather than more compact objects such as galaxy clusters. 

Despite the large order of magnitude discrepancy between our modeled and 
observed integrated
cross-correlation functions, we are consistent with several other studies,
including \citet{BM97, WI98, NW00} and \citet{BSM01}, because they have
also noted such large discrepancies.
A number of potential causes for the
mismatches have been discussed in the literature.  On small angular scales
less than a few arcminutes, correlations can be explained by the halo model,
which indicates that the strength of the signal is very dependent on the halo
occupation properties of various types of galaxies.  Depending on the specific
types of galaxies used, the halo model predicts an enhancement of a factor of 2
- 10 on small angular scales relative to nonlinear clustering models by
\citet{B95} and \citet{DB97} \citep{JSS03}.  There is still a discrepancy on
the $\gtrsim 30\arcmin $ scales within this study, however, and a number of
other phenomena have been investigated to explain this.  
\citet{W00}, for example,
investigated extending lensing formalism by including higher order general
relativistic terms, but this results only in a $\sim 10\%
$ increase in the theoretical cross-correlation estimate.  Many other studies
have suggested dust, but extragalactic dust would tend to reduce 
cross-correlation, but there is probably little or no dust in clusters 
\citep{NWM03}.  Galactic dust could create positive correlations, but one
would also expect to see positive correlations between faint stars and QSOs,
which we do not detect.  Studies
involving radio-selected QSOs (see Table 1), which should be unaffected by 
dust, are also inconsistent with significant 
amounts of Galactic obscuration.  We have also compared various
cross-correlation estimators.  Each are dependent to varying degrees on
selection effects and window functions, but each yield very similar results. 
This likely implies that at least the APM and Hamburg-ESO catalogues are fair
samples of their respective populations and that systematic effects are not
inflating the cross-correlation estimates.  

So in order to further investigate the discrepancy between the model and 
the observations,
we have also measured the slope of the QSO number-magnitude counts, $\alpha_m$,
as a function of redshift using the same sample of QSOs for each redshift bin
that was described in Section~\ref{ccz} (the binwidth is, again, 
$\Delta z = 0.4$).  
Figure~\ref{alpha} shows that $\alpha_m$ is indeed above the critical value of
0.4 for all redshifts, with perhaps a slight overall increase in the slope with
increasing QSO redshift. Typical values are on the order of $\alpha_m \approx
0.6$.   The error bars represent the $1\sigma$ RMS variation in the slope
of $\Delta z = 0.1$ subsets of the $\Delta z = 0.4$ bins of our QSO sample
and uncertainties at redshifts $z > 1.5$ are
underestimated due to low QSO numbers.

\begin{figure}
\epsscale{0.8}
\plotone{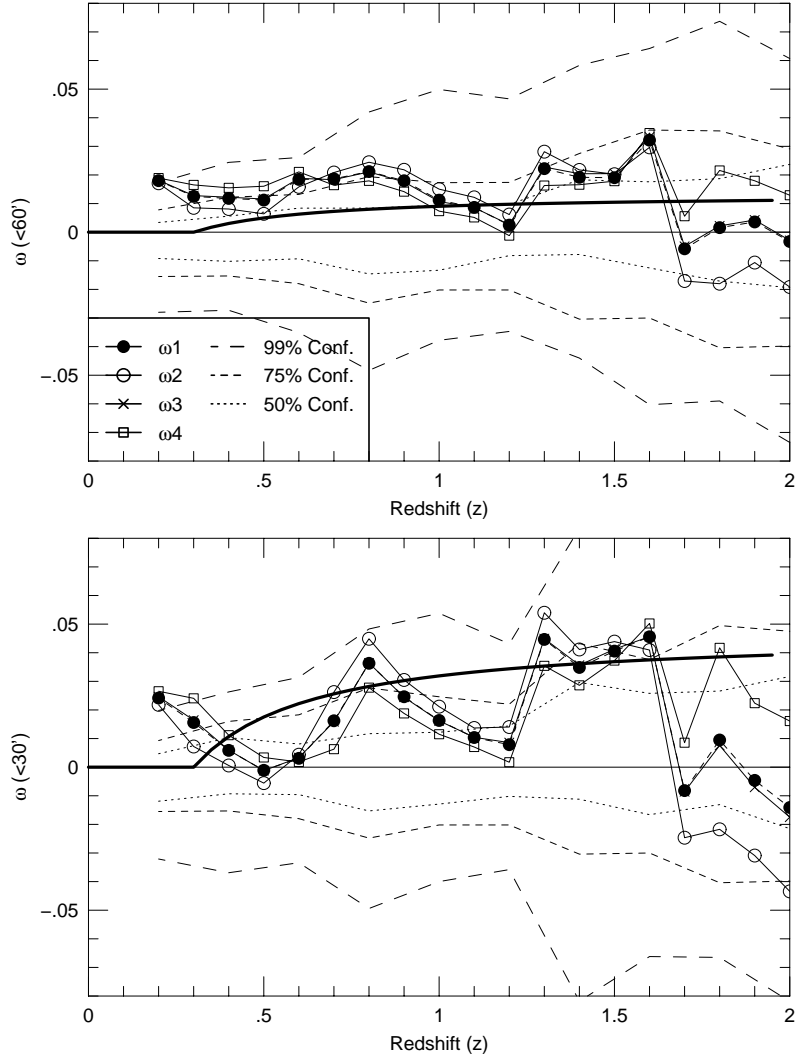}
\epsscale{1.0}
\caption{Measured and modeled integrated correlation functions vs. redshift 
with confidence
limits.  Each of the four correlation estimators from the Methods section are
shown.  {\it Top:} Correlation functions integrated over separations $\leq
60\arcmin.$ {\it Bottom:} The same, but now integrated over separations $\leq
30\arcmin.$  
The bold dark line in each case represents a large scale structure
lensing model with a linear CDM mass power spectrum, following \citet{BS01}.
Dotted, dashed and long-dashed lines represent $50$, $75$ and $99\%$ 
confidence limits from 1000 iterations of our monte carlo error estimation.
Note the positive signal detected at redshifts $z > 0.25$ and that  
$\omega_{3}$ nearly coincides with $\omega_{1}$ on both panels.}
\label{xvz}

\end{figure}

\begin{figure}
\epsscale{0.8}
\plotone{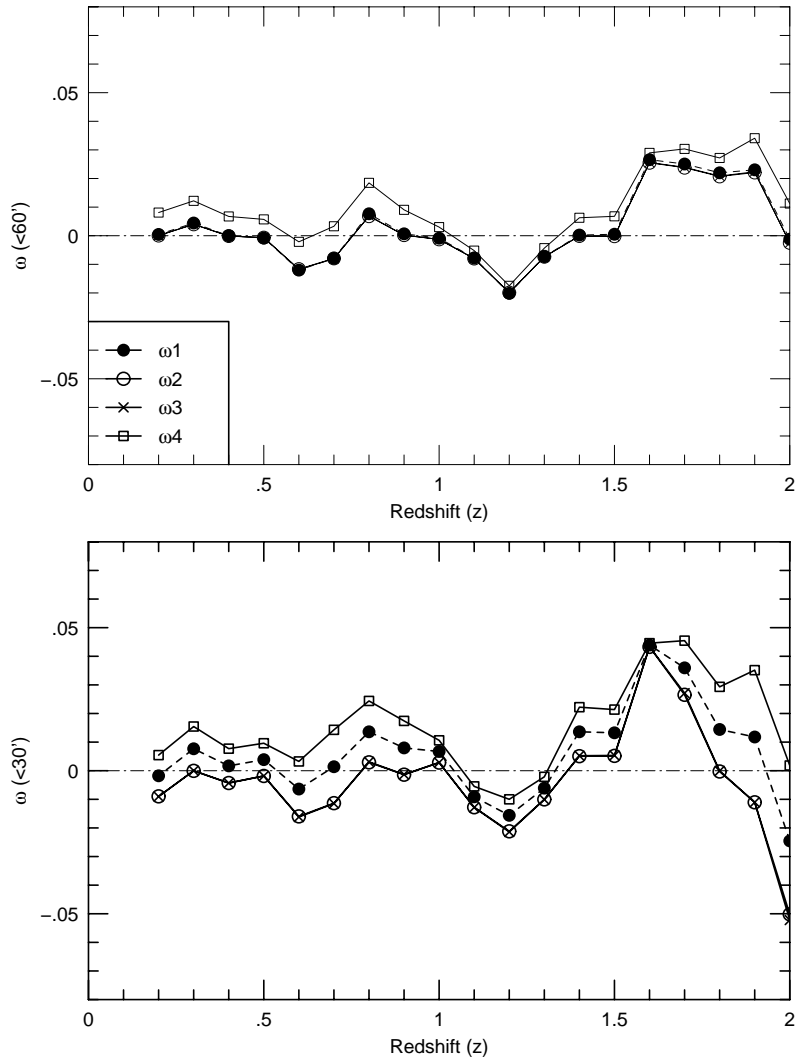}
\epsscale{1.0}
\caption{Same as Figure~\ref{xvz} only substituting galaxies 
with a selection of APM stars with magnitudes
between $16 \leq R \leq 17.5$. One would expect no 
correlation between stars and distant QSOs.
Unlike the positive signal seen in
Figure~\ref{xvz}, we indeed find no correlation between stars and QSOs.  
Offsets between the various
estimators are due to differing shapes of the individual correlation functions
from each estimator.  Some contamination of the stellar sample by galaxies is
possible.  Several estimators are in close agreement for the $60\arcmin$
correlations while $\omega_2$ and $\omega_4$ are most similar in $30\arcmin$ 
correlations.} 
\label{xvzs}
\end{figure}

\begin{figure}
\epsscale{0.8}
\plotone{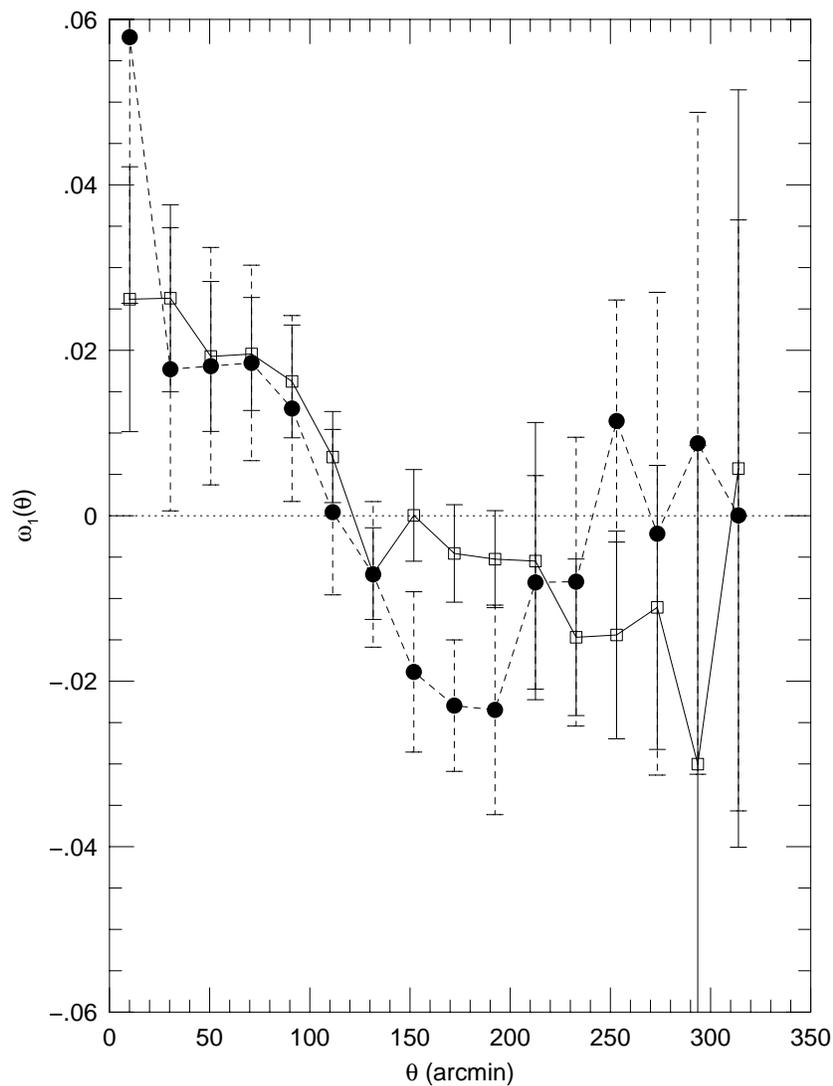}
\epsscale{1.0}
\caption{Correlation functions from two redshift cuts.  {\it Empty Squares:} All 113 QSOs
with $0 < z < 0.2$ (physical associations).  {\it Solidd dots:} The 79 QSOs with 
$0.6 < z < 1.0,$ corresponding to the location of the strongest signal in the
integrated correlations from Figure~\ref{xvz} (magnification bias).  Despite their different physical cause, both
correlation functions appear to be very similar considering the statistical
uncertainties.}
\label{cz}
\end{figure}

\begin{figure}
\epsscale{0.8}
\plotone{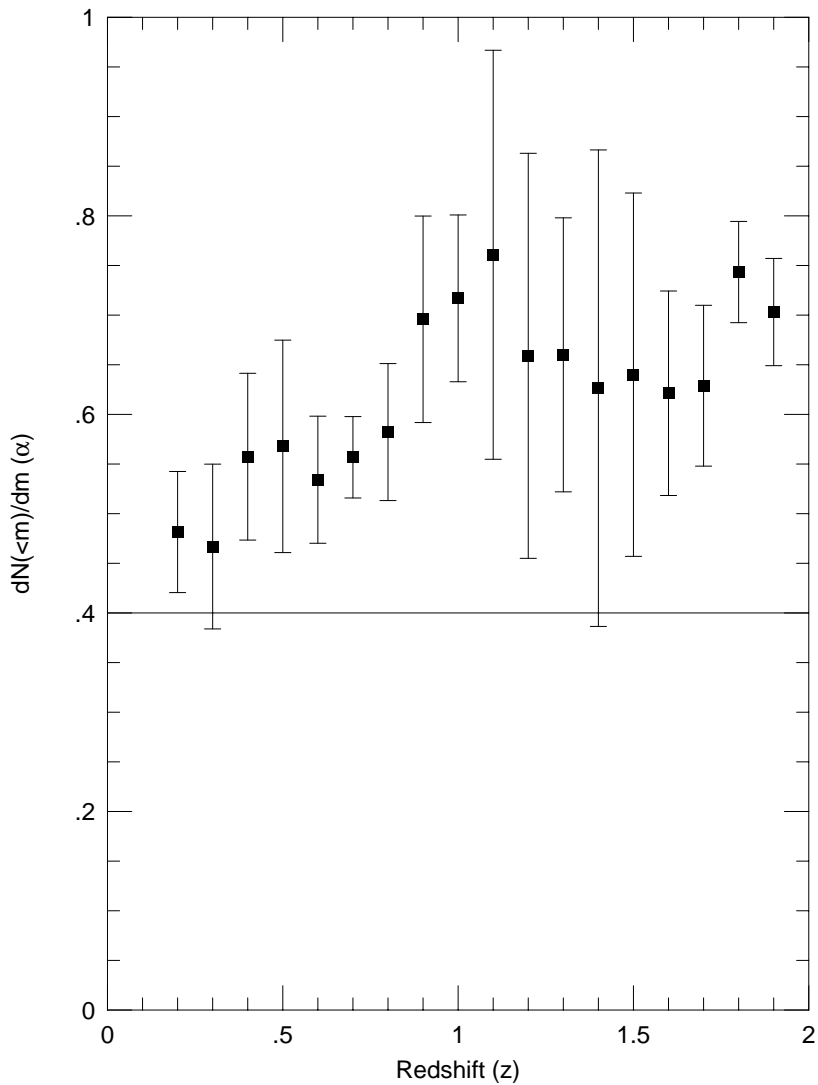}
\epsscale{1.0}
\caption{Average slope of the QSO number-magnitude counts plotted vs.redshift. 
Redshift bins have width $\Delta z = 0.4$ and each bin contains the same QSOs as
the bins found in Figure~\ref{xvz}. 
The value of the slope of the QSOs within the Hamburg-ESO Catalogue is
$\alpha_{m} > 0.4$ for all redshifts, implying the expectation of positive
correlations due to magnification bias.  Error bars represent the $1\sigma$
RMS of slopes from subsets of the data.  Uncertainties given at redshifts
between $1.5 \leq z \leq 2.0$ are artificially low due to low QSO numbers per
bin.}
\label{alpha}
\end{figure}

\begin{figure}
\epsscale{0.8}
\plotone{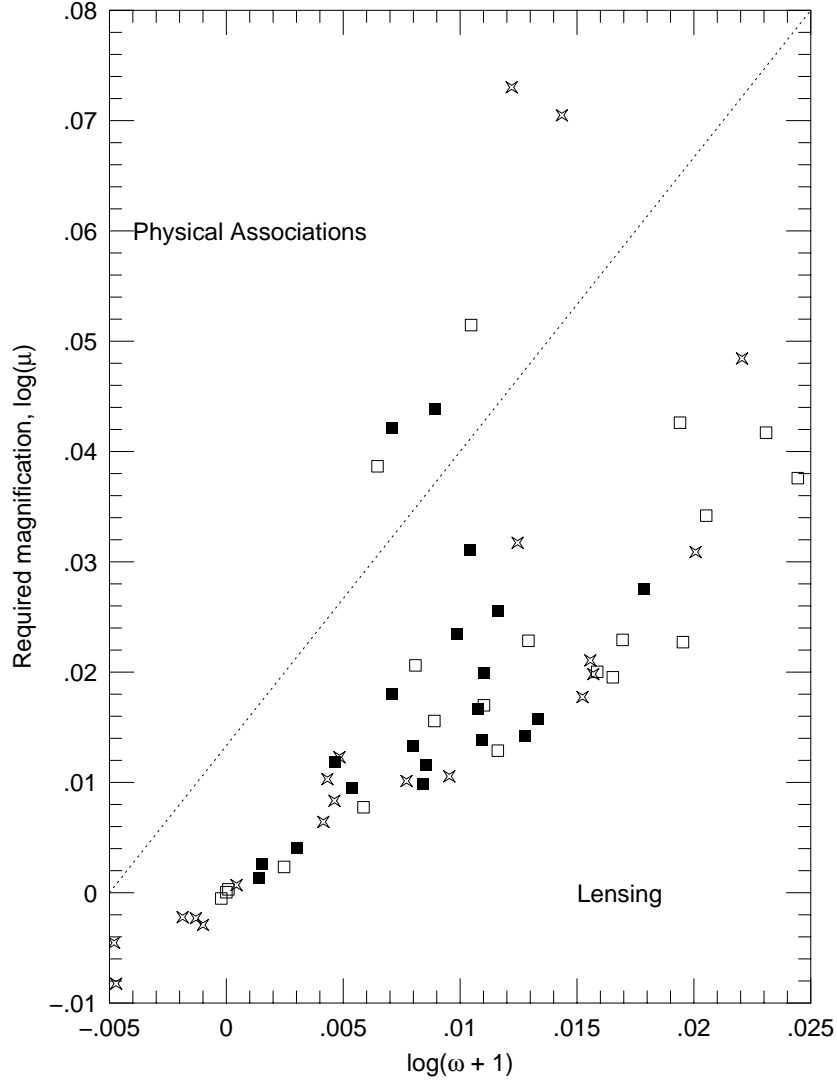}
\epsscale{1.0}
\caption{The required magnification to generate the measured
integrated correlation,assuming lensing and given the measured slope of the QSO 
number-magnitude counts, $\alpha_m(z)$ for each redshift bin from Figure 
\ref{xvz}. X's, empty squares and filled squares correspond to 
points from $\omega(<15\arcmin), \omega(<30\arcmin)$ and $\omega(<60\arcmin)$,
respectively.
The points can be separated between the two lowest redshift bins,
and the rest.  (The dotted line separating the two populations is, outside the
fact that it divides the two populations, arbitrary.) 
The requirement to see such large
magnifications to account for the measured correlations when 
compared to the rest of the points implies a different
mechanism is the cause of the measured correlations:  Physical associations.
}
\label{mag}
\end{figure}

Using the measured $\alpha_m$ with the measured correlation strengths, we
determined the degree of magnification present, assuming that the signal is due
to magnification bias (see Equation~\ref{lens}).  
The magnification is plotted vs. measured correlation
strength in Figure~\ref{mag}, using the same data found in Figures~\ref{xvz} and
\ref{alpha}.  Because the physical associations do not obey Equation~\ref{lens},
the estimated magnification should not necessarily compare with that expected
from the magnification bias affect points.   Therefore, we are able to separate
those points corresponding
to low-redshift physical associations from the lensed points at higher 
redshift with this plot.

Finally, because of the rather large amplitude discrepancy with the models, we
investigated whether the observed overdensities could be due to extinction by
Galactic dust.  As mentioned in Section~\ref{gd}, there is no correlation
between the measured galaxy overdensities about the QSOs and 
the differences in extinction level between the QSO lines of sight and the
field.  If dust were the cause, one would have
difficulty in explaining the observed underdensities that have been measured in
other studies that employ faint QSOs, as expected from magnification bias.
Patchy Galactic dust can only produce positive cross-correlations, not
anti-correlations.  Both of these lines of evidence suggest 
that Galactic dust is not the cause of the large amplitude
in the cross-correlation signal.  Furthermore considering that the signal
appears to behave as expected, qualitatively, 
for magnification bias, it seems likely that this is
the cause of the observed galaxy-QSO cross-correlations.  

Yet we are still left with the long-standing problem of the discrepancy with
models.  There remain a number of avenues to explore in hopes of solving this
problem.  Dust, whether Galactic or extragalactic in origin, are unlikely to
skew the values of the cross-correlations significantly.  Galactic dust is
incapable of explaining anti-correlations and extragalactic dust can not result
in positive correlations.  Both types of correlations are observed in accordance
to magnification bias.  We will investigate the effect of catalogue masks and
selection effects in \citet{NW05}, though there are still a number of other 
places to look for a solution.  For example, perhaps models need to account for
multiple or nonlinear deflections for a given light ray as it travels from
source to observer.  There has also been a suggestion by Bassett and Kunz
that since galaxy-QSO cross-correlations are one of two types of studies 
(the other involving Type-Ia supernovae) in which both the 
luminosity and angular diameter distances are sampled, there may be an asymmetry
between the two distances \citep{BK03, BK04}.  This asymmetry could in principle
be caused by essentially any process that would hinder a photon from reaching 
an observer, including Compton scattering off the ionized IGM or something 
much more exotic such as photon decay.

\acknowledgements

We thank Lutz Wisotzki for his provision of the initial release of the 
Hamburg-ESO QSO catalogue and for his very useful comments regarding the 
preparation of the survey and his equally helpful suggestions for
this paper.

\clearpage

\renewcommand{\baselinestretch}{0.75}
\setlength{\tabcolsep}{0mm}
\setlength{\footnotesep}{2mm}
{\scriptsize
\begin{deluxetable}{lccccccc}
\rotate
\tablewidth{0pt}
\tablecaption{\label{tbl:priorwork} Previous Results from Galaxy-QSD Association}
\startdata
\hline\hline\\ [-1.5ex] & {\scriptsize Galaxy} 
 & {\scriptsize Galaxies} & {\scriptsize QSO} & 
 {\scriptsize QSOs} & {\scriptsize QSO} & 
 {\scriptsize Angular} &  \\
{\scriptsize Study }& {\scriptsize Source} & {\scriptsize Mag/Flux} & 
{\scriptsize Source} & {\scriptsize Mag/Flux} & 
{\scriptsize Redshifts} & {\scriptsize Scales} & 
{\scriptsize Detected} \\ 
\hline \\ [-1.5ex]

	{\scriptsize \citet{T86}} & {\scriptsize \citet{T86}} & 
	{\scriptsize $R \lesssim 21$} & {\scriptsize \citet{HB80}} &
	{\scriptsize $17 < V < 20$} & {\scriptsize $0.13 - 0.48$} & 
	{\scriptsize $< 30\arcsec$}  & {\scriptsize $+$} \\
	& {\scriptsize "} & {\scriptsize "} & {\scriptsize "} &
	{\scriptsize "} & {\scriptsize $0.96 - 1.5$} & 
	{\scriptsize "} & {\scriptsize $+$ }\\
	{\scriptsize \citet{w88}} & {\scriptsize APM UKST} & {\scriptsize $B_{J}
	< 21$} & {\scriptsize \citet{w88}} & {\scriptsize } & 
	{\scriptsize $1 - 3$} & {\scriptsize $\lesssim 10\arcsec$} &
	{\scriptsize $+$} \\
	{\scriptsize \citet{F88}} & {\scriptsize \citet{F88}} & 
	{\scriptsize $r < 21\fm5$} & {\scriptsize 1 Jy Flat spectrum QSOs} &
	{\scriptsize  $S_{5GHz} \geq 1$Jy} & {\scriptsize $z > 1.7$ } & 
	{\scriptsize $< 60\arcsec$} & {\scriptsize $+$} \\
	{\scriptsize \citet{F90}} & {\scriptsize Lick Catalogue} & 
	{\scriptsize $m_{pg} < 18\fm9$} & {\scriptsize 1 Jy Flat spectrum QSOs} &
	{\scriptsize $S_{5GHz} \geq 1$Jy} & 
	{\scriptsize $z \geq 1$} & {\scriptsize $\sim 10\arcmin$} & 
	{\scriptsize $+$} \\
	& {\scriptsize  "} & {\scriptsize "} & {\scriptsize "} &
	{\scriptsize "} & {\scriptsize $0.2 - 1$} & 
	{\scriptsize "} & {\scriptsize $0$} \\
	{\scriptsize \citet{HLF90}} & {\scriptsize \citet{HLF90}} & 
	{\scriptsize $R \leq 21$} & {\scriptsize 3CR radio galaxies} &
	{\scriptsize  $S_{178MHz}: 4.5 - 9$Jy} & {\scriptsize $z \geq 1$} & 
	{\scriptsize $\sim 5\arcsec$} & {\scriptsize $+$} \\
	{\scriptsize \citet{HRV91}} & {\scriptsize \citet{HRV91}} & 
	{\scriptsize $R < 21$} & {\scriptsize \citet{VV89}} &
	{\scriptsize $17 < V < 20$} & {\scriptsize $0.9 - 1.5$} & 
	{\scriptsize $\sim 15\arcsec$} & {\scriptsize $+$} \\
	{\scriptsize \citet{d92}} & {\scriptsize \citet{d92}} & 
	{\scriptsize $V < 18\fm0$} & {\scriptsize \citet{d92}} &
	{\scriptsize  $16 \leq V \leq 18$} & {\scriptsize $0.7 - 2.5$}
	& {\scriptsize $15\arcsec$} & {\scriptsize $+$} \\
	{\scriptsize \citet{BS94}} & {\scriptsize IRAS Faint Source Cat.} & 
	{\scriptsize $S_{60\mu m}:0.3 - 1$Jy} & {\scriptsize 1 Jy Flat spectrum
	QSOs} & {\scriptsize $S_{5GHz} 
	\geq 1$Jy} & {\scriptsize $z \gtrsim 1.25$} & 
	{\scriptsize $\sim 10\arcmin$} & {\scriptsize $+$} \\
	\hline
	{\scriptsize \citet{BFS88}} & {\scriptsize APM UKST} &{\scriptsize N/A} & 
	{\scriptsize UVX} &
	{\scriptsize $ B < 20\fm0$} & {\scriptsize $< 2.2$} & 
	{\scriptsize $< 10\arcmin$} & {\scriptsize $-$} \\
	{\scriptsize \citet{RWH94}} & {\scriptsize Zwickey Clusters} & 
	{\scriptsize N/A} & {\scriptsize LBQS} &
	{\scriptsize $B \leq 18\fm5$} & {\scriptsize $1.4 - 2.2$} &
	 {\scriptsize $\lesssim 140\arcmin$} & {\scriptsize $+$} \\
	{\scriptsize \citet{BM97}} & {\scriptsize COSMOS/UKST} &
	{\scriptsize $B_{J} \lesssim 21$} & {\scriptsize LBQS} &
	{\scriptsize $B_{J} < 20.5$} &
	{\scriptsize $0.3 - 4$} & {\scriptsize $< 15\arcmin$} & 
	{\scriptsize $-$} \\
	{\scriptsize \citet{WI98}} & {\scriptsize APM UKST} &
	{\scriptsize $16 \leq R \leq 20.46$} & {\scriptsize LBQS} & 
	{\scriptsize $ B_{J} \leq
	18\fm0$} & {\scriptsize $< 100\arcmin$} & 
	{\scriptsize $< 3.3$} & {\scriptsize $+$} \\
	{\scriptsize \citet{CS99}} & {\scriptsize COSMOS/UKST} & 
	{\scriptsize $B_J < 20.5$} & {\scriptsize UVX} 
	& {\scriptsize $17.9 < B < 20.65$} & {\scriptsize $0.3
	- 2.2$} & {\scriptsize $< 10\arcmin$} & {\scriptsize $-$} \\
	{\scriptsize \citet{G03}} & {\scriptsize SDSS} &
	{\scriptsize $r' = 19 - 22$} & {\scriptsize SDSS} &
	{\scriptsize $i' < 19.1$ }& 
	{\scriptsize $0.8 - 2.5$} & {\scriptsize $< 10\arcmin$} & 
	{\scriptsize $+$}\\
	{\scriptsize \citet{m03}} & {\scriptsize APM and SDSS} & 
	{\scriptsize $B < 20.5$} & {\scriptsize 2dF}
	& {\scriptsize $B < 20.5$} & {\scriptsize $< 3$} &
	{\scriptsize $\lesssim 10\arcmin$} & {\scriptsize $-$}\\
	\hline
	{\scriptsize \citet{SS95}} & {\scriptsize Zwickey Clusters} & 
	{\scriptsize N/A} & {\scriptsize 1 Jy Catalogue} &
	{\scriptsize $S_{5GHz} > 1$Jy} & {\scriptsize $0 - 0.5$} &
	{\scriptsize $\lesssim 15\arcmin$} & {\scriptsize $+$} \\
	{\scriptsize \citet{SS95}} & {\scriptsize Zwickey Clusters} & 
	{\scriptsize N/A} & {\scriptsize 1 Jy Catalogue} &
	{\scriptsize $S_{5GHz} > 1$Jy} & {\scriptsize $0.5 - 1.5$} &
	{\scriptsize $\lesssim 15\arcmin$} & {\scriptsize $-$} \\
	{\scriptsize \citet{BM97}} & {\scriptsize COSMOS/UKST} &
	{\scriptsize $B_{J} \lesssim 21$} & {\scriptsize Parks catalogue} & 
	{\scriptsize $S_{11cm} > 0.5$Jy}
	& {\scriptsize $0.3 - 4$} & {\scriptsize $< 15\arcmin$} & 
	{\scriptsize $+$} \\
	{\scriptsize \citet{NW00}} & {\scriptsize APM UKST} &  
	{\scriptsize $18.5 < R < 20$} & {\scriptsize 1 Jy Catalogue} & 
	{\scriptsize $S_{5GHz} > 1$Jy} & 
	{\scriptsize $> 0.5$} & {\scriptsize $< 60\arcmin$} & 
	{\scriptsize $+$} \\
	{\scriptsize \citet{BSM01}} & {\scriptsize COSMOS/UKST} & 
	{\scriptsize $B_{J} < 20.5$} & {\scriptsize 1 Jy Catalogue} 
	 & {\scriptsize $S_{5GHz} > 1$Jy} & {\scriptsize $<3$} & 
	 {\scriptsize $< 15\arcmin$ } & {\scriptsize $+$} \\
	{\scriptsize \citet{BSM01}} & {\scriptsize COSMOS/UKST} &
	{\scriptsize $B_{J} < 20.5$} & {\scriptsize 0.5 Jy Catalogue} & 
	{\scriptsize $S_{5GHz} > 0.5$Jy} & 
	{\scriptsize $<3$} & {\scriptsize $< 15\arcmin$} & {\scriptsize $+$} \\
\hline \\ [-1.75ex]
\enddata
\end{deluxetable}}

\clearpage

\renewcommand{\baselinestretch}{1}
\setlength{\tabcolsep}{2mm}
{\small
\begin{table}[t] \caption[2D Estimator dependence on catalogue characteristics]
{\label{estcat2} 2D Estimator dependence on catalogue characteristics}
\begin{center}
\begin{tabular}{|c|c|c|}
\hline
	{\bf Correlator} & {\bf Pair counts} & {\bf Dependence} \\
\hline\hline
	\multicolumn{3}{|c|}{{\bf Autocorrelation estimators}} \\ \hline
	 & & \\ 
	\raisebox{1.5ex}[0pt]{$\omega_{GG1}$} & 
	\raisebox{1.5ex}[0pt]{$\frac{DD}{\langle DR \rangle} - 1$} & 
	\raisebox{1.5ex}[0pt]{$\frac{\hat{\omega}_{GG}
	 + \phi_{G} - \bar{\sigma}_{G} - \phi_{G}\bar{\sigma}_{G}}{(1 +
	 \phi_{G})(1 + \bar{\sigma}_{G})}$ }\\ \hline
	 & & \\
	\raisebox{1.5ex}[0pt]{$\omega_{GG2}$} & 
	\raisebox{1.5ex}[0pt]{$\frac{DD \cdot \langle RR \rangle}
	{\langle DR \rangle^{2}} - 1$} & 
	\raisebox{1.5ex}[0pt]{$\frac{\hat{\omega}_{GG} - \phi_{G}^{2}}{(1 +
	 \phi_{G})^{2}}$ }\\ \hline
	 & & \\
	\raisebox{1.5ex}[0pt]{$\omega_{GG3}$} & 
	\raisebox{1.5ex}[0pt]{$\frac{DD -2\langle DR \rangle + \langle RR 
	\rangle}{\langle RR \rangle}$} & 
	\raisebox{1.5ex}[0pt]{$\frac{\hat{\omega}_{GG} - 2\phi_{G}\bar{\sigma}_{G}
	+ \bar{\sigma}_{G}^{2}}{(1 +
	 \bar{\sigma}_{G})^{2}}$ }\\ \hline
	  & & \\
	\raisebox{1.5ex}[0pt]{$\omega_{GG4}$} & 
	\raisebox{1.5ex}[0pt]{$\frac{DD}{\langle RR \rangle} - 1$}  & 
	\raisebox{1.5ex}[0pt]{$\frac{\hat{\omega}_{GG} + 2\phi_{G} - 
	2\bar{\sigma}_{G}- \bar{\sigma}_{G}^{2}}{(1 +
	 \bar{\sigma}_{G})^{2}}$ }\\ \hline 
	  & & \\
	\raisebox{1.5ex}[0pt]{$\omega_{GG5}$} & 
	\raisebox{1.5ex}[0pt]{$\frac{\langle DR \rangle}{\langle RR \rangle} 
	- 1$}  & 
	\raisebox{1.5ex}[0pt]{$\frac{\phi_{G} - \bar{\sigma}_{G}}{(1 +
	 \bar{\sigma}_{G})}$ }\\ \hline 
	  \hline
	\multicolumn{3}{|c|}{{\bf Cross-correlation estimators}} \\ \hline
	 & & \\ 
	\raisebox{1.5ex}[0pt]{$\omega_{1}$} & 
	\raisebox{1.5ex}[0pt]{$\frac{D_{G}D_{Q}}{\langle D_{G}R_{Q} \rangle} - 
	1$} & 
	\raisebox{1.5ex}[0pt]{$\frac{\hat{\omega}_{GQ}
	 + \phi_{Q} - \bar{\sigma}_{Q}(1+\phi_{G})}{(1 +
	 \phi_{G})(1 + \bar{\sigma}_{Q})}$ }\\ \hline
	 & & \\
	\raisebox{1.5ex}[0pt]{$\omega_{2}$} & 
	\raisebox{1.5ex}[0pt]{$\frac{D_{G}D_{Q} \cdot \langle R_{G}R_{Q} 
	\rangle}{\langle D_{G}R_{Q} \rangle\langle R_{G}D_{Q} \rangle} - 1$} & 	
	\raisebox{1.5ex}[0pt]{$\frac{\hat{\omega}_{GQ} - \phi_{G}\phi_{Q}}{(1 +
	 \phi_{G})(1 + \phi_{Q})}$ }\\ \hline
	 & & \\
	\raisebox{1.5ex}[0pt]{$\omega_{3}$} & 
	\raisebox{1.5ex}[0pt]{$\frac{D_{G}D_{Q} - \langle D_{G}R_{Q} \rangle 
	- \langle R_{G}D_{Q} \rangle + \langle R_{G}R_{Q} 
	\rangle}{\langle R_{G}R_{Q} \rangle}$} & 
	\raisebox{1.5ex}[0pt]{$\frac{\hat{\omega}_{GQ} - \phi_{G}\bar{\sigma}_{Q}
	-\phi_{Q}\bar{\sigma}_{G} + \bar{\sigma}_{G}\bar{\sigma}_{Q}}{(1 +
	 \bar{\sigma}_{G})(1 + \bar{\sigma}_{Q})}$ }\\ \hline
	  & & \\
	\raisebox{1.5ex}[0pt]{$\omega_{4}$} & 
	\raisebox{1.5ex}[0pt]{$\frac{D_{G}D_{Q}}{\langle R_{G}R_{Q} \rangle} 
	- 1$}  & 
	\raisebox{1.5ex}[0pt]{$\frac{\hat{\omega}_{GQ} + \phi_{G} - 
	 \bar{\sigma}_{G} + \phi_{Q} - \bar{\sigma}_{Q} +\bar{\sigma}_{G}
	 \bar{\sigma}_{Q}}{(1 +
	 \bar{\sigma}_{G})(1+ \bar{\sigma}_{Q})}$ }\\ \hline 
	  & & \\
	\raisebox{1.5ex}[0pt]{$\omega_{5}$} & 
	\raisebox{1.5ex}[0pt]{$\frac{\langle D_{G}R_{Q} \rangle}
	{\langle R_{G}R_{Q} \rangle} - 1$}  & 
	\raisebox{1.5ex}[0pt]{$\frac{\phi_{G} - \bar{\sigma}_{G}}{(1 +
	 \bar{\sigma}_{G})}$ }\\ \hline 
	  & & \\
	\raisebox{1.5ex}[0pt]{$\omega_{6}$} & 
	\raisebox{1.5ex}[0pt]{$\frac{\langle R_{G}D_{Q} \rangle}
	{\langle R_{G}R_{Q} \rangle} - 1$}  & 
	\raisebox{1.5ex}[0pt]{$\frac{\phi_{Q} - \bar{\sigma}_{Q}}{(1 +
	 \bar{\sigma}_{Q})}$ }\\ \hline 
	  & & \\
	\raisebox{1.5ex}[0pt]{$\omega_{7}$} & 
	\raisebox{1.5ex}[0pt]{$\frac{D_{G}D_{Q}}
	{\langle R_{G}D_{Q} \rangle} - 1$}  & 
	\raisebox{1.5ex}[0pt]{$\frac{\hat{\omega}_{GQ} + \phi_{G} -
	\bar{\sigma}_{G}(1 + \phi_{Q})}{(1 +
	 \bar{\sigma}_{G})(1 + \phi_{Q})}$ }\\ 
	 \hline 
\end{tabular}
\end{center}
\end{table}}

\clearpage

\clearpage

\clearpage

\clearpage

\clearpage

\clearpage

\clearpage

\clearpage

\clearpage

\end{document}